\documentclass[reprint,superscriptaddress,prl,showpacs]{revtex4-1}

\usepackage{graphicx}
\usepackage{epstopdf}
\usepackage[ansinew]{inputenc}
\usepackage{array}
\usepackage{color}
\usepackage{amsmath}
\usepackage{amsxtra}
\usepackage{amstext}
\usepackage{amssymb}
\usepackage{latexsym}
\usepackage{float}
\usepackage{soul}

\begin{document}
\title{Local Spectroscopic Characterization of Spin and Layer Polarization in WSe$_2$}
\author{Matthew Yankowitz}
\affiliation{Physics Department, University of Arizona, 1118 E 4th Street, Tucson, AZ 85721, USA}
\author{Devin McKenzie}
\affiliation{Physics Department, University of Arizona, 1118 E 4th Street, Tucson, AZ 85721, USA}
\author{Brian J. LeRoy}
\email{leroy@physics.arizona.edu}
\affiliation{Physics Department, University of Arizona, 1118 E 4th Street, Tucson, AZ 85721, USA}
\date{\today}


\begin{abstract}

We report scanning tunneling microscopy (STM) and spectroscopy (STS) measurements of monolayer and bilayer WSe$_2$. We measure a band gap of 2.21 $\pm$ 0.08 eV in monolayer WSe$_2$, which is much larger than the energy of the photoluminescence peak indicating a large excitonic binding energy. We additionally observe significant electronic scattering arising from atomic-scale defects. Using Fourier transform STS (FT-STS), we map the energy versus momentum dispersion relations for monolayer and bilayer WSe$_2$. Further, by tracking allowed and forbidden scattering channels as a function of energy we infer the spin texture of both the conduction and valence bands. We observe a large spin-splitting of the valence band due to strong spin-orbit coupling, and additionally observe spin-valley-layer coupling in the conduction band of bilayer WSe$_2$.  

\end{abstract}

\maketitle

Transition metal dichalcogenides (TMDs) have gained great popularity recently for their potential applications in optical and electronic devices~\cite{Xu2014}. The band gaps of these materials lie in the visible part of the spectrum, and undergo an indirect-to-direct transition at monolayer thickness.  In monolayers, the band extrema (valleys) lie at the K and K' points~\cite{comment}, and the lack of inversion symmetry leaves these valleys physically distinguishable (for example, via optical selection rules). The heavy transition metals in TMDs also contribute to strong spin-orbit coupling (SOC) for band edge carriers, with especially large effects expected for the tungsten dichalcogenides such as WSe$_2$. This SOC leads to an out-of-plane Zeeman-type spin splitting of the valence band as large as 500 meV, with opposite spin polarizations at the K and K' valleys due to time reversal symmetry leading to a coupling of the spin and valley degrees of freedom. 

AB-stacked bilayer TMDs are inversion symmetric, and as a consequence the spin and valley polarizations of the top layer are opposite those of the bottom, thus effectively restoring spin and valley degeneracy. They additionally exhibit an extra degree of freedom characterized by a layer pseudospin. In WSe$_2$, the SOC is larger than the interlayer hopping amplitude, thus localizing the carriers by layer near the Brillouin zone (BZ) corners and leading to a coupling of the spin, valley, and layer degrees of freedom (see Ref.~\onlinecite{Xu2014} for a review of these properties). Numerous spintronic and valleytronic applications have been proposed to exploit these degrees of freedom in TMD monolayers and bilayers~\cite{Xiao2012,Cao2012,Mak2012,Lu2013,Wu2013,Gong2013,Yuan2013,Jones2013,Jones2014,Xu2014,Mak2014,Yuan2014}.
   
%%%%%%%%%%%%%%%
\begin{figure}[t]
\includegraphics[width=8.6cm]{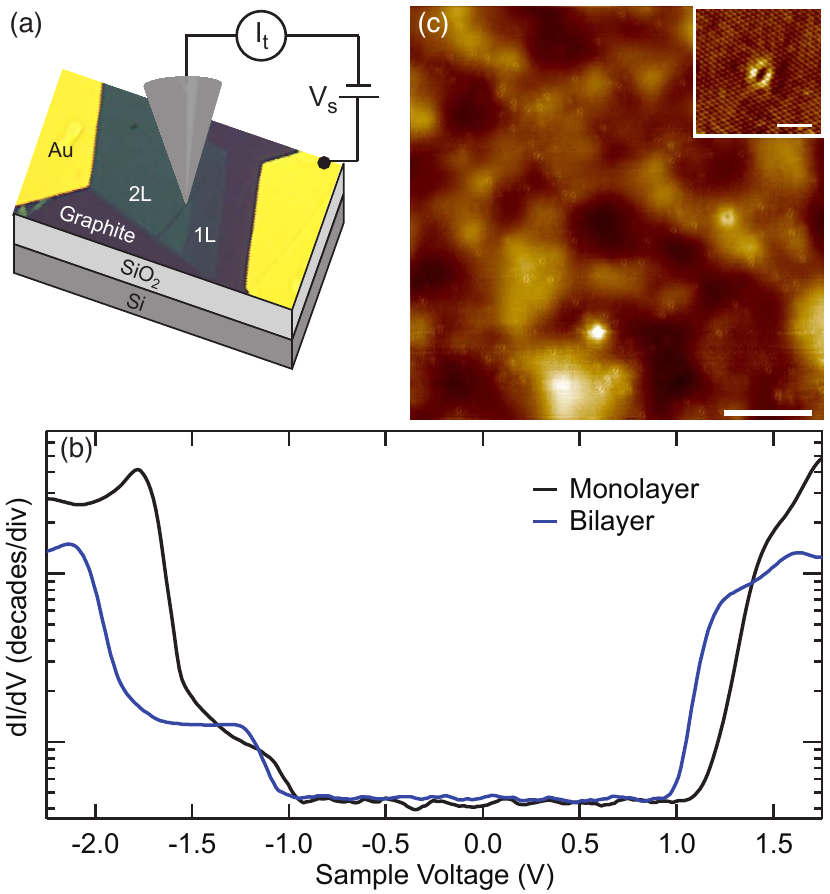} 
\caption{(a) Schematic of the experimental setup showing the STM tip and an optical microscope image of the measured WSe$_2$ on graphite sample. (b) Typical dI/dV spectroscopy of monolayer and bilayer WSe$_2$ on a log scale. (c) STM topography of a typical region of bilayer WSe$_2$. There are two distinct defect types, with the large defects far less prevalent than the small defects. The scale bar is 20 nm. Inset: Atomic resolution of a small defect. The scale bar is 2 nm.}
\newpage
\label{fig:schematic}
\end{figure}
%%%%%%%%%%%%%%%%%%

Significant work has been performed to characterize these polarizations in WSe$_2$ with direct optical excitation~\cite{Jones2013,Jones2014,Aivazian2015,Srivastava2015} and spin-resolved ARPES~\cite{Riley2014,Le2015}. However, the former is unable to capture physics away from direct transitions (i.e. away from the BZ corners) and the latter cannot probe unoccupied (conduction band) states. Here we directly probe the electronic states of monolayer and bilayer WSe$_2$ via STM and STS measurements in order to gain a deeper understanding of its band structure and internal quantum degrees of freedom. Fig.~\ref{fig:schematic}(a) shows a schematic of our measurement setup. All measurements were performed in ultrahigh vacuum at a temperature of 4.5 K. WSe$_2$ sits on a graphite flake to provide a conducting substrate for collecting the tunnel current. The sample was fabricated using a dry-peel transfer technique~\cite{Wang2013,Zomer2014}. WSe$_2$ and graphite were mechanically exfoliated onto separate SiO$_2$ substrates, then the WSe$_2$ flake was picked up using a polycarbonate film and micro-mechanically aligned onto a graphite flake. Cr/Au electrodes were written using electron beam lithography, and the sample was annealed overnight in vacuum at 300 $^{\circ}$C. 

Fig.~\ref{fig:schematic}(b) shows representative dI/dV spectroscopy for monolayer and bilayer WSe$_2$. We extract the electronic band gap of both by measuring the energy range where the dI/dV spectroscopy is zero, from which we find E$_g$ = 2.21 $\pm$ 0.08 eV for monolayer and 2.14 $\pm$ 0.05 eV for bilayer WSe$_2$. This is significantly larger than the typically observed optical gaps in monolayer and bilayer WSe$_2$ ($\sim$1.65 eV and 1.6 eV, respectively)~\cite{Zhao2013,Zeng2013,Zhang2014}, indicative of a giant excitonic binding energy~\cite{Ugeda2014}. We note that this spectroscopically extracted gap actually represents the separation of the onset of tunneling into the valence and conduction band extrema, which, due to the finite width of the bands, is smaller than the separation between the center of the valence band maximum and the conduction band minimum. The peaks and/or inflection points in the dI/dV spectrum likely represent the band centers, which would imply a band gap a few hundred meV larger than that extracted above, in reasonably good agreement with the GW level DFT calculated gap~\cite{Ding2011,Ramasubramaniam2012}. Due to small changes in the tip work function resulting from the tip shaping procedure (for example, pulsing a gold particle onto the end of the tip), we have occasionally observed small ($<$150 meV) rigid shifts of the entire dI/dV spectrum in sample voltage. To account for this, we take dI/dV spectroscopy before all measurements, and when necessary offset the sample voltage to match the dI/dV reference spectra of Fig.~\ref{fig:schematic}(b).

\begin{figure}[t]
\includegraphics[width=8.6cm]{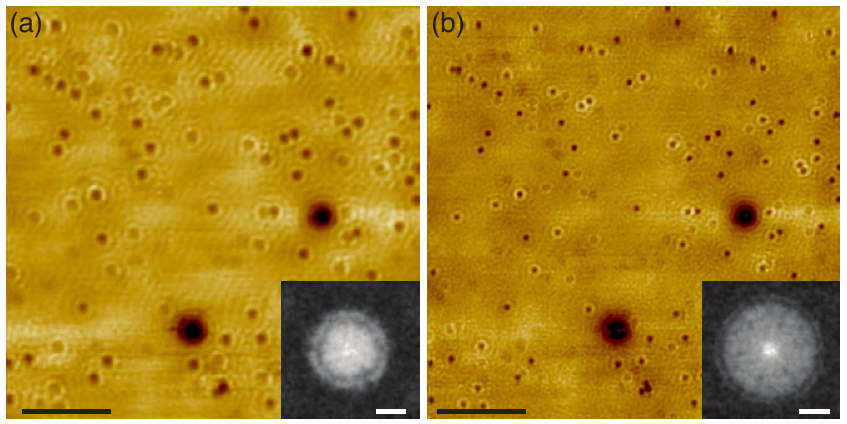} 
\caption{dI/dV maps at constant sample biases of (a) -1.35~V and (b) -1.5~V on bilayer WSe$_2$ taken over the same area as Fig.~\ref{fig:schematic}(c). The two species of defects (large and small) are easily distinguishable. The small defects additionally appear either strong or weak, depending on whether they are in the top or bottom layer. Friedel oscillations surround the defects, and their wavelength grows shorter as the energy moves deeper into the valence band.  Scale bar is 20 nm for both. Insets: Fourier transforms of the maps. As the scattering wavelength grows shorter, the size of the scattering disk in momentum space grows larger. Scale bar is 3 nm$^{-1}$ for both.}
\newpage
\label{fig:defects}
\end{figure}

Topographic maps of WSe$_2$ exhibit a significant density of defects. Fig.~\ref{fig:schematic}(c) shows one such example for bilayer WSe$_2$, in which two species of defects are present. There are small atomic-scale defects throughout the image (the inset is an atomically-resolved image of one such defect), along with two larger scale defects. These defects are more clearly visible in spatially resolved dI/dV maps, as shown in Figs.~\ref{fig:defects}(a) and (b) for the same area as in Fig.~\ref{fig:schematic}(c), taken at sample voltages of V$_s$ = -1.35~V and -1.5~V, respectively. In bilayer WSe$_2$, defects of a given species appear with two clearly distinct strengths, where the weaker defects presumably reside in the lower WSe$_2$ layer. This indicates we are capable of probing electrons in both layers. In monolayer WSe$_2$, all defects are observed with similar strength as they all reside in the same layer~\cite{SM}. By counting the number of defects in numerous dI/dV maps taken on monolayer WSe$_2$, we find a defect density of 1.1 $\pm$ 0.3 x 10$^{12}$ cm$^{-2}$ for the smaller defects, and approximately two orders of magnitude lower density for the large defects. Given the similar appearance of sulfur vacancies observed in MoS$_2$~\cite{Lu2014} it is likely that the small defects we observe here are due to selenium vacancies, although a full defect characterization is outside the scope of this work.

%%%%%%%%%%%%%%%
%%%%%%%%%%%%%%%%%%

\begin{figure}[t]
\includegraphics[width=8.6cm]{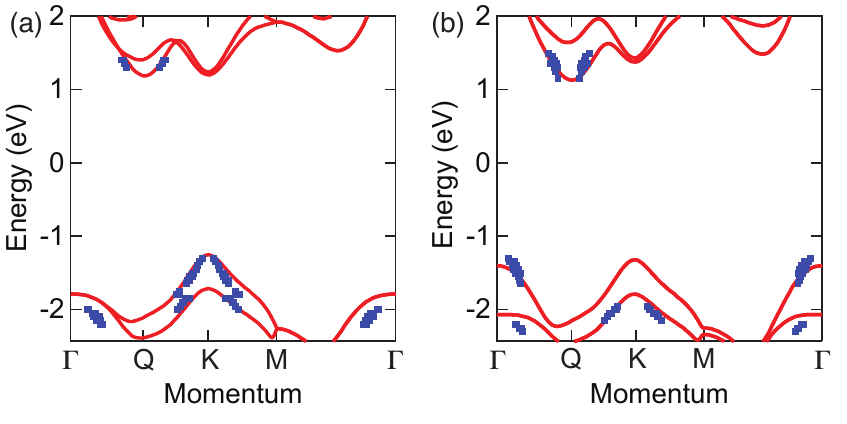} 
\caption{Band structure of (a) monolayer and (b) bilayer WSe$_2$. The solid red lines are \textit{ab initio} calculations of the band structure. The calculated band gap is manually set to match the spectroscopy as described in the main text. Blue dots are the experimentally extracted momentum. Points are assigned to bands as described in the supplementary material. All points have an uncertainty in momentum of less than 10\% of their pre-offset value (and typically less than 5\%), arising from the uncertainty in identifying the size of the scattering disk in the Fourier transforms. There is an energy smearing between 5 and 20 mV on all points due to the ac voltage added to acquire the dI/dV signal. These have been omitted for clarity.}
\newpage
\label{fig:dispersion}
\end{figure}

These defects act as scattering centers for the electrons in WSe$_2$, resulting in Friedel oscillations visible as concentric rings surrounding each defect~\cite{Friedel1952}. The wavelength of these oscillations at a given energy is controlled by the band structure of the WSe$_2$. So, for example, since the valence band grows wider as the energy is lowered (Fig.~\ref{fig:defects}(a) to (b)), the wavelength of the oscillations grows shorter as they are inversely proportional. FT-STS provides direct visualization of the available scattering channels for electrons in momentum space~\cite{Hoffman2002}. The long-wavelength Friedel oscillations of Figs.~\ref{fig:defects}(a) and (b) are due to intravalley scattering processes and show up as disk-like features at the center of their respective Fourier transforms (insets). The signal in the Fourier transform can be understood as a map of the joint density of states (JDOS) of the band structure. The intravalley scattering process with the largest possible momentum transfer connects one side of a given valley to the other, setting the radius of the disk observed in the Fourier transform. The disk is then filled in by all other smaller momentum transfer scattering processes connected within a single valley. As a result, the radius of the central scattering disk gives a measure of the size of the Fermi surface at a given sample voltage. The blue dots in Figs.~\ref{fig:dispersion}(a) and (b) mark the momentum extracted from the size of these scattering disks as a function of sample voltage, providing experimental maps of the band structures of monolayer and bilayer WSe$_2$ (see the Supplemental Material~\cite{SM} for details on wave vector extraction and band assignments). Data from different sets of dI/dV maps taken with different tips and at different spots on the sample are all plotted together, as they follow the same general trend. Data points have been plotted on both sides of a given valley since the Fermi surfaces are roughly circular.

%%%%%%%%%%%%%%%
%%%%%%%%%%%%%%%%%%

We confirm our extracted energy versus momentum dispersion by comparing with the calculated \textit{ab initio} band structures of monolayer and bilayer WSe$_2$ with spin-orbit coupling terms included (solid red lines in Fig.~\ref{fig:dispersion}), taken directly from Ref.~\onlinecite{Zeng2013}. Since the size of the band gap is not well constrained in these calculations, it has been manually set to match the band edges extracted from the dI/dV spectroscopy in Fig.~\ref{fig:schematic}(b) (plus a small offset to account for the finite band width). Our experimental results are in reasonable agreement with the theoretically anticipated bands; most notably the effective masses generally match well with the theoretically predicted values. The extracted bands at $\Gamma$ near -2 eV are around a hundred meV below the calculated bands for both monolayer and bilayer WSe$_2$, suggesting possible discrepancies between the DFT calculation and the true band structure. Disagreements may also result from mixing of the scattering signals of two bands.

%%%%%%%%%%%%%%%
\begin{figure*}[t]
\includegraphics[width=17.8cm]{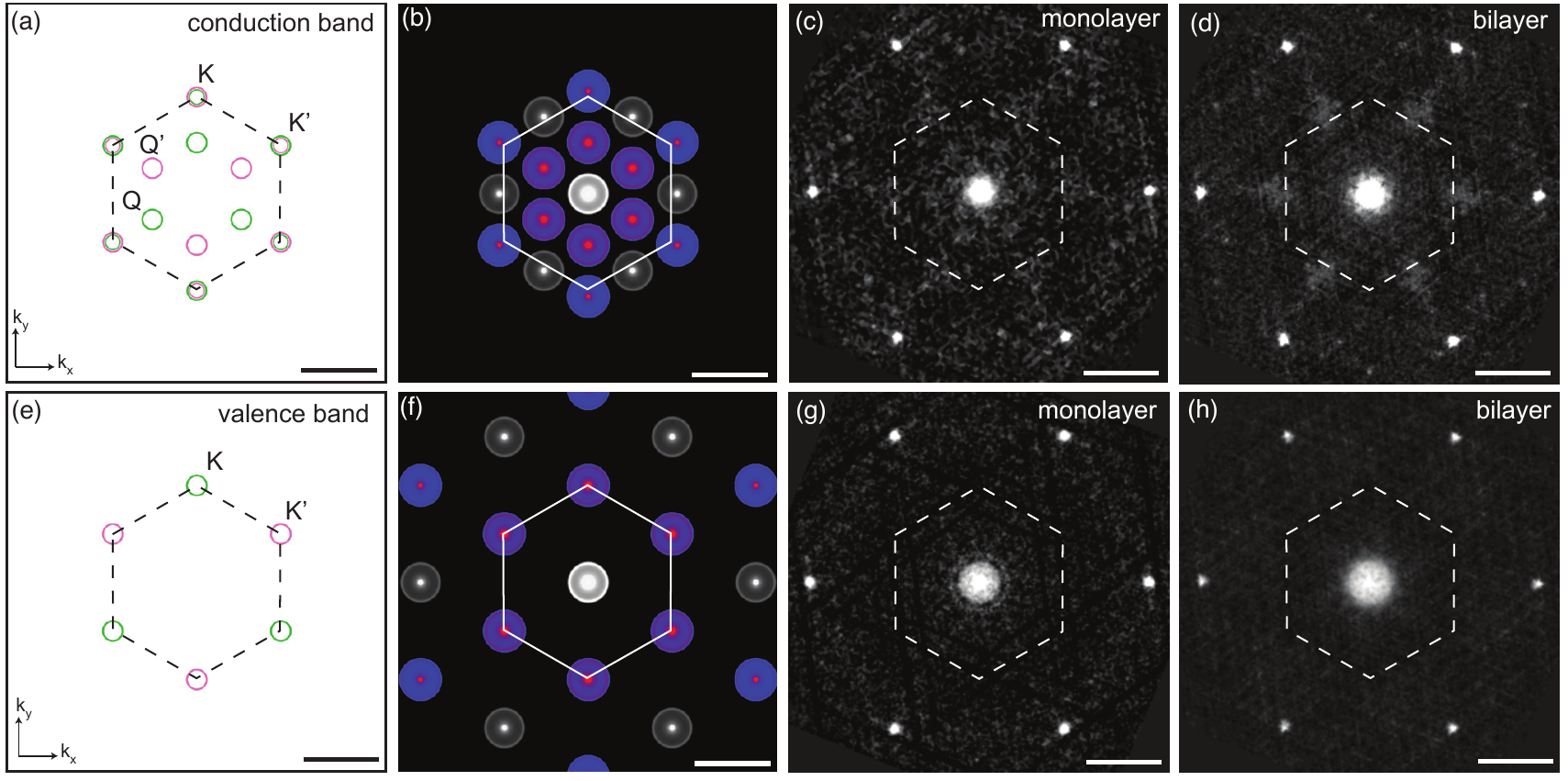} 
\caption{(a)  General schematic of the conduction band of monolayer and bilayer WSe$_2$. The relative sizes of the bands depends on the layer number. In bilayer WSe$_2$ only the Q-point bands are present at the very bottom of the conduction band, and the K-point bands are spin-degenerate. The green denotes spin up bands and the pink denotes spin down.  (b) Simulated JDOS of the band structure shown in (a). Only the Q-point bands are considered. Scattering resonances shown in white are always allowed, but colored features are only possible with no spin texture on the bands. (c) Experimental FT-STS on monolayer WSe$_2$ at V$_s$ = +1.5 V. The sharp resonances in a hexagon around the outside of the image are due to the atomic lattice. Only scattering allowed with the spin-textured bands is present (white features), implying the anticipated spin texture of the Q-point bands. (d) Experimental FT-STS on bilayer WSe$_2$ at V$_s$ = +1.3 V. The same subset of scattering features allowed in monolayer WSe$_2$ are observed, indicating strong layer polarization in bilayer WSe$_2$. (e) Schematic of the band structure of monolayer WSe$_2$ between the spin-split valence bands. (f) Simulated JDOS of the band structure shown in (e). (g) Experimental FT-STS on monolayer WSe$_2$ at V$_s$ = -1.6 V. No intervalley scattering resonances are observed (blue features in (f)), implying the anticipated spin texture of the K-point bands. (h) Experimental FT-STS on bilayer WSe$_2$ at V$_s$ = -2.0 V. The intervalley scattering is also suppressed, similarly indicating strong layer polarization in bilayer WSe$_2$. The scale bars are 10 nm$^{-1}$ for all, and the white hexagons represent the first BZ.}
\newpage
\label{fig:scattering}
\end{figure*}
%%%%%%%%%%%%%%%%%%

In addition to the intravalley scattering channels, there could, in principle, be scattering between different valleys. Such scattering occurs with larger momentum transfer than intravalley scattering, and therefore the resonances should be further from the center of the Fourier transforms. Due to strong spin-orbit coupling in WSe$_2$, the valence band at the K point, which is primarily comprised of $m$=$\pm$2 $d$-orbitals of W~\cite{Roldan2014a,Roldan2014b}, is spin-split by around 450 meV, with bands having opposite spins in the K and K' valleys. Fig.~\ref{fig:scattering}(e) shows a schematic of the band structure of monolayer WSe$_2$ at an energy between the two spin-split valence bands. There are six Fermi pockets surrounding the K and K' valleys, alternating between spin up at K (green) and spin down at K' (pink). Without this spin texture, atomic-scale defects would induce intervalley scattering, identified by resonances in the Fourier transform at $4\pi/3a_0$, where $a_0$ is the WSe$_2$ lattice constant ($\sqrt{3}$ times closer to the center than the lattice resonances, and rotated from them by 30$^{\circ}$). However, the spin polarization of the valleys is expected to suppress these resonances, as the intervalley scattering process now additionally requires a spin-flip which non-magnetic defects are unable to provide. 

We simulate the Fourier transform anticipated for each case by calculating the both the spin-dependent and -independent JDOS of the band structure shown in Fig.~\ref{fig:scattering}(e)~\cite{Roushan2009}. The result is plotted in Fig.~\ref{fig:scattering}(f), where the colored features correspond to scattering resonances expected only in the case of bands lacking spin polarization. The central resonance represents the intravalley scattering discussed previously, and its diameter is twice the diameter of the K-point bands. Due to the geometry of the chords which can connect two points on the constant energy contour, the scattering signal decays away from the center of the Fourier transform but becomes large again just around the circumference of the disk, which is in good agreement with the features we observe in the maps shown in the insets of Fig.~\ref{fig:defects}. Fig.~\ref{fig:scattering}(g) shows a comparable symmetrized Fourier transform of an atomically resolved dI/dV map taken on monolayer WSe$_2$~\cite{SM}. The features anticipated from K-valley mixing are conspicuously absent despite the high density of atomic-scale defects normally capable of inducing such scattering, implying the anticipated spin-polarization of the valleys. In contrast, this valley mixing is readily observable in graphene with atomic-scale defects~\cite{Rutter2007}, which lacks the spin polarization of the K-point valleys~\cite{CastroNeto2009}.

In the conduction band, the lowest energy bands exist around the K and Q points, the latter of which is not a high symmetry point but occurs about halfway between $\Gamma$ and K. Fig.~\ref{fig:scattering}(a) shows a general schematic of the Fermi surface in the conduction band along with the spin polarization theoretically anticipated for WSe$_2$~\cite{Kormanyos2015}. While there is only a small spin-splitting anticipated at the K point ($<$50 meV) since these bands are predominately $m$=0 $d$-orbitals of W~\cite{Kosmider2013}, the Q-point bands are expected to be split by over 200 meV in monolayer and 500 meV in bilayer WSe$_2$ since they are comprised of both $m$=$\pm$1 $p$-orbitals of Se and $m$=$\pm$2 $d$-orbitals of W~\cite{Roldan2014a,Roldan2014b}. Fig.~\ref{fig:scattering}(b) shows the JDOS calculation for the conduction band with the same color coding scheme as in Fig.~\ref{fig:scattering}(f). The K-point bands are removed from the calculation to reflect the fact that they are not expected to factor significantly into the observed scattering processes due to the reduced tunnel probability into K in the monolayer~\cite{SM}, and do not exist for the bottom-most $\sim$250 meV of the bilayer conduction band. Fig.~\ref{fig:scattering}(c) shows the Fourier transform at the bottom of the conduction band of monolayer WSe$_2$. Similar to the case of the valence band, only the features allowed assuming spin-polarized bands are present (note that the features corresponding to Q to Q scattering along the sides of the hexagonal Brillouin zone are present but very weak.)

The spin degeneracy of the spin-split bands is restored in bilayer WSe$_2$ due to the inversion symmetry of the layers. So, for example, if a given band in the bottom layer is spin down, that same band in the upper layer will be spin up. Despite the restored spin degeneracy, the Fourier transforms of our experimental dI/dV maps in the bilayer conduction band (Fig.~\ref{fig:scattering}(d)) and the valence band (Fig.~\ref{fig:scattering}(h)) do not exhibit all the resonances anticipated in the simulation~\cite{tunnel}. They do, however, match the exact subset of resonances expected given spin-polarized bands (white features in Figs.~\ref{fig:scattering}(b) and (e)). Put another way, the bilayer maps exhibit only the features observed in the monolayer maps, implying an effective spin-polarization of the bilayer WSe$_2$ bands. This spin-polarization arises from strong layer-polarization of the bilayer WSe$_2$. Since there is only a small component of $p_z$ in the Q- and K-point bands, the spin-orbit coupling dominates over the interlayer hopping, thus enhancing the layer polarization. As we are capable of probing electrons in both layers of bilayer WSe$_2$, the observation of nearly identical Fourier transforms for monolayer and bilayer WSe$_2$ provides direct evidence of strong spin-layer-valley entanglement in both the valence and conduction bands.

We have presented local tunnel spectroscopy measurements of monolayer and bilayer WSe$_2$. By examining electronic scattering from atomic-scale defects we are able to map out portions of the band structure of the material, as well as to identify scattering pathways forbidden by spin and layer polarizations of the bands. Our results directly demonstrate that intervalley electronic scattering is suppressed in the valence band of monolayer WSe$_2$ as a result of the large spin-splitting of those bands. Furthermore, we observe strong layer-polarization in bilayer WSe$_2$, providing the first experimental evidence of the spin-valley-layer coupling in the conduction band. While previous applications have been proposed to utilize this effect for slightly hole-doped bilayer WSe$_2$~\cite{Gong2013}, our results present a path towards realizing these novel combined spintronic and valleytronic applications in devices of either doping.

\begin{acknowledgments}

The authors thank Gui-Bin Liu and Wang Yao for supplying the calculated band structures and for valuable discussions. The authors also thank Allan H. MacDonald for valuable theoretical discussions. The work at Arizona was partially supported by the U. S. Army Research Laboratory and the U. S. Army Research Office under contract/grant number W911NF-14-1-0653 and the National Science Foundation DMR-0953784.

\end{acknowledgments}

\bibliographystyle{apsrev4-1}
\bibliography{references}

\begin{thebibliography}{37}
\bibitem [{com()}]{comment}%
  \BibitemOpen
  \href@noop {} {}\bibinfo {note} {In WSe$_2$, some theoretical and
  experimental results suggest the lowest conduction band exists at the Q
  point, although it is very nearly degenerate with the band at K.}\BibitemShut
  {Stop}%
\bibitem [{tun()}]{tunnel}%
  \BibitemOpen
  \href@noop {} {}\bibinfo {note} {The bilayer valence band energy shown is
  within the lower of the spin-split bands at K, where we have the highest
  tunnel sensitivity into the Brillouin zone corners. The schematic in Fig.
  4(e) also represents the most relevant bands in bilayer WSe$_2$ at this
  energy.}\BibitemShut {Stop}%
\end{thebibliography}

\end{document}